\documentstyle[graphicx,prl,aps,multicol,epsf]{revtex}

\begin{document}
\draft

\title{Many-body diagrammatic expansion in a Kohn-Sham basis:\\
implications for Time-Dependent Density Functional Theory of excited states}
\author{I.~V.~Tokatly\cite{MIET} and O.~Pankratov}
\address{Lehrstuhl f\"ur Theoretische Festk\"orperphysik, 
Universit\"at Erlangen-N\"urnberg, Staudtstr. 7/B2, 91054 Erlangen,
Germany}
\date{\today} 
\maketitle

\begin{abstract}
We formulate diagrammatic rules for many-body perturbation theory
which uses Kohn-Sham (KS) Green's functions as basic propagators. The
diagram technique allows to study the properties of the dynamic
nonlocal exchange-correlation (xc) kernel $f_{xc}$. We show that the
spatial non-locality of $f_{xc}$ is strongly frequency-dependent. In
particular, in extended systems the non-locality range diverges at the
excitation energies. This divergency is related to the discontinuity
of the xc potential.
\end{abstract}
 

\begin{multicols}{2}

Time-dependent density functional theory (TDDFT) \cite{RG} offers
a possibility to extend the powerful density-functional formalism
\cite{DG} to excited states of many-body systems \cite{Petersilka1,Burke}. 
A substantial improvement of excitation energies with respect to 
KS eigenvalues was obtained for atoms and molecules \cite{Petersilka1,Burke,Petersilka2,Molecules}
using a variety of approximations for a dynamic xc kernel $f_{xc}=
\delta v_{xc}({\bf r},t)/\delta n({\bf r'},t')$ ($v_{xc}$ is a xc 
potential). However in solids the wrong KS band gap remains unchanged
regardless the approximation used, albeit the dielectric function 
is on average improved \cite{sem}.

This situation, as we show below, reflects an extremely nonlocal
behavior of $f_{xc}$ at excitation frequencies. The non-locality range
is as large as the system size and hence diverges in extended systems.
None of the up-to-date approximations account for this behavior as they 
all employ the adiabatic (frequency-independent) xc kernels.

In this paper we develop a perturbative technique with KS Green's 
functions as the bare 
propagators. In essence, it is a diagrammatic expansion of 
Sham-Schl\"uter equation \cite{Sham}, which maintains a correct electron density 
in every order of the perturbation theory. 
We find that at resonant frequencies the kernel $f_{xc}$ is proportional 
to the discontinuity of $v_{xc}$. This explains the anomalous non-locality 
of $f_{xc}$, since a constant shift of a potential due to an extra particle 
is felt by another particle anywhere in the system.

In the framework of TDDFT the excitation energies are commonly 
calculated \cite{Petersilka1,Burke,Petersilka2} from the poles 
of the linear response function $\chi({\bf r},{\bf r'},\omega)$. The latter is
related to the KS susceptibility $\chi_{S}({\bf r},{\bf r'},\omega)$
by 
\begin{equation}
{\hat \chi}(\omega) = {\hat \chi}_{S}(\omega) + {\hat \chi}_{S}(\omega)\left[
{\hat V}_{C}+{\hat f}_{xc}(\omega)\right]{\hat \chi}(\omega), 
\label{1}
\end{equation}
where $V_{C}=e^{2}/|{\bf r}-{\bf r'}|$ is a bare Coulomb
repulsion and the kernel $f_{xc}$
enters as an additional dynamic interaction.

Alternatively, the poles of $\chi({\bf r},{\bf r'},\omega)$ can be found as the eigenvalues of a
linearized equation for  density matrix
\begin{eqnarray}\nonumber 
& &\left[\omega - {\hat H}_{S}({\bf r}_{1})+ {\hat H}_{S}({\bf r}_{2})\right] 
\delta \rho({\bf r}_{1},{\bf r}_{2})\\ 
 &-& \rho_{S}({\bf r}_{1},{\bf r}_{2})
\int d{\bf r}\left[{\tilde V}_{\omega}({\bf r}_{1},{\bf r}) - 
{\tilde V}_{\omega}({\bf r},{\bf r}_{2})\right]
\delta \rho({\bf r},{\bf r}) = 0 
\label{2}
\end{eqnarray}
where ${\tilde V}_{\omega}=V_{C}+f_{xc}(\omega)$,
${\hat H}_{S}({\bf r})$ is the KS Hamiltonian, and 
$\rho_{S}({\bf r},{\bf r}')=
\sum_{j}n_{j}\psi^{*}_{j}({\bf r})\psi_{j}({\bf r}')$ is the
equilibrium KS density matrix with the KS orbitals 
$\psi^{*}_{j}({\bf r})$.
Equation (\ref{2}) clearly shows that a correction to the KS
excitation energies originates from the 
Hartree-type energy of the excitation-induced density fluctuation 
$\delta n({\bf r})= \delta \rho({\bf r},{\bf r})$.  
In the KS basis equation (\ref{2}) takes the form
\begin{equation}
(\omega - \omega^{S}_{ij})\delta \rho_{ij} - \gamma_{ij}
\sum_{kl}\langle \Phi_{ij}|
{\tilde V}_{\omega}
|\Phi_{kl}\rangle
\delta \rho_{kl}=0,
\label{3}
\end{equation}
where 
$\omega^{S}_{ij}=E^{S}_{i}-E^{S}_{j}$ is a KS excitation energy, 
$\gamma_{ij}=n_{i}-n_{j}$ is the difference of the
occupation numbers and
$\Phi_{ij}({\bf r})=\psi^{*}_{i}({\bf r})\psi_{j}({\bf r})$.
The ordinary perturbation theory gives the energy shift
$\Delta \omega_{ij} =\omega_{ij}-\omega^{S}_{ij}$
in the first order as
\begin{equation}
\Delta\omega^{(1)}_{ij}= 
\langle\Phi_{ij}({\bf r})|
V_{C}({\bf r},{\bf r}')+ f_{xc}({\bf r},{\bf r}',\omega^{S}_{ij})
|\Phi_{ij}({\bf r}')\rangle,
\label{5}
\end{equation}
which is identical to the result of the first-order
Laurent expansion of $\chi({\bf r},{\bf r'},\omega)$
\cite{Petersilka1}. The perturbation theory easily 
allows to obtain the higher-order corrections as well. It is worth noting 
that equations (\ref{2}) and (\ref{3}) and the perturbative 
result (\ref{5}) are equally valid for finite and for extended systems.  

Let us consider the dependence of the first-order correction (\ref{5}) 
on the size of a system $L\sim V^{1/3}$ at fixed average density $N/V$. As  
$\Phi_{ij}({\bf r})$ contains a normalization factor $1/V$, the first term in Eq. (\ref{5}) 
is proportional to $e^{2}/L$. This is the Coulomb energy of the density variation 
due to an electron-hole excitation, which is infinitesimally small in
extended systems. The second term crucially depends on the non-locality
of $f_{xc}$ i.e. on the extension of a xc-hole. At $\omega = 0$ the non-locality range 
is about the interparticle distance $l\sim (V/N)^{1/3}$. This feature is reproduced by 
the popular OEP approximation \cite{Petersilka1} whereas in adiabatic 
LDA \cite{ALDA} $f_{xc}$ is a point interaction. Assuming that  
at resonance frequency $f_{xc}(\omega_{ij})$ has a similar
non-locality we find that the second term in Eq. (\ref{5}) is proportional
to $1/N$ and vanishes as $L^{-3}$ (see, however, \cite{tail}). Thus {\it any} xc-kernel which
is finite and decays at infinity does not contribute to $\omega_{ij}$ in extended systems. Using the 
many-body perturbative approach formulated below, we show that a non-vanishing xc 
correction arises due to the divergence
of  the non-locality scale of  $f_{xc}(\omega_{ij})$. 

\begin{figure}[t]
\centering
\includegraphics[
width=0.25\textwidth]{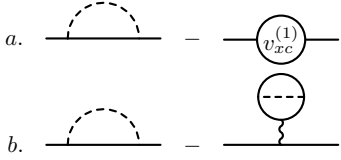}
\caption{First-order corrections to the Green's function.}
\label{fig1}
\end{figure}
\noindent
We employ the Matsubara formalism at nonzero temperature $T$ which
enables us to obtain any physical retarded function through analytic continuation \cite{LandauIX}. 
Assuming that the xc-potential $v_{xc}({\bf r})$ and the ground state density
are known, we represent the Hamiltonian of a system as a
sum of the KS Hamiltonian ${\hat H}_{S}$ and the 
perturbation ${\hat V}$, where 
\begin{equation}
{\hat V}={\hat W}' - \sum_{k=1}^{\infty}\int d{\bf r} 
v^{(k)}_{xc}({\bf r}){\hat n({\bf r})}.
\label{6}
\end{equation}
In Eq. (\ref{6}) ${\hat W}'$ is a two-particle
interaction with the Hartree part being subtracted, ${\hat n({\bf r})}$
is a density operator and we
assume that the xc-potential $v_{xc}$ can be expanded in a power series
$v_{xc}=\sum_{k=1}^{\infty}v^{(k)}_{xc}$, where $v^{(k)}_{xc}\sim
e^{2k}$. 

Following the standard procedure \cite{LandauIX} we define the Green's
function 
\begin{equation}
G(X,X')= - 
\langle T_{\tau}\Psi_{S}(X)\Psi^{+}_{S}(X'){\hat \sigma}\rangle/
\langle {\hat \sigma}\rangle,
\label{7}
\end{equation}
where $X=({\bf r},\tau)$ ($\tau$ is an imaginary time), $\Psi_{S}(X)$
is a field operator in a KS interaction representation and 
${\hat\sigma}$ is a Matsubara S-matrix \cite{LandauIX} which corresponds 
to the perturbation ${\hat V}$ of Eq. (\ref{6}). 
The angular brackets in Eq. (\ref{7}) denote averaging over the KS
equilibrium state. 

A formal graphical expansion of $G(X,X')$ 
contains along with the pair interaction terms the 
diagrams related to the scattering by the ``external'' 
potentials $v_{xc}^{(k)}$. To achieve a closed scheme one needs 
a complementary graphical representation of $v_{xc}({\bf r})$, which can be
obtained from the condition of density conservation .
The Green's function (\ref{7}) can be written in the form
$G=G_{S}+\sum_{k=1}^{\infty}G^{(k)}$, where $G_{S}$ is a KS Green's 
function and $G^{(k)}$ is a $k$-th order correction. 
As the KS system possesses an exact density, the variation 
of the density due to the interaction (\ref{6}) must
vanish. Applying this requirement to every order 
we arrive at the system of equations:
\begin{equation}
\delta n^{(k)}({\bf r})=T\sum_{n=-\infty}^{\infty}
G^{(k)}({\bf r},{\bf r},\omega_{n}) = 0,  
\label{9}
\end{equation}
\begin{figure}[t]
\centering
\includegraphics[
width=0.3\textwidth]{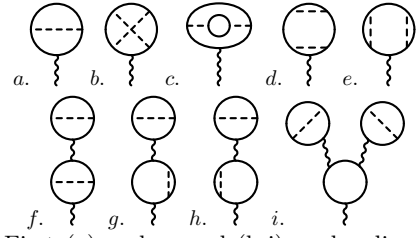}
\caption{First (a) and second (b-i) -order diagrams for xc potential.}
\label{fig2}
\end{figure}
\noindent
where $\omega_{n}=\pi T(2n+1)$.
This system is equivalent to well-known Sham-Schl\"uter equation
\cite{Sham} (see also \cite{Sham-Farid}).  
A successive solution of Eq. (\ref{9}) allows 
to construct $v^{(k)}_{xc}$ for every $k$. For example, the first-order
correction $G^{(1)}$ is presented 
in Fig. 1a, where the solid line is the KS
Green's function and dashed line is the Coulomb
interaction.  Substituting $G^{(1)}$ into 
Eq. (\ref{9}) we get  $v^{(1)}_{xc}$ as shown in Fig. 2a, where the wiggled 
line stands for the inverse KS response function 
$\chi^{-1}_{S}({\bf r},{\bf r}')$. The final
first-order correction to the Green's function is shown in Fig. 1b. 
Note that $v^{(1)}_{xc}$ (Fig. 2a) exactly corresponds to 
the x-only OEP $v_{x}$.
Given $v^{(1)}_{xc}$ we solve Eq. (\ref{9}) for $k=2$ and obtain eight
graphs for $v^{(2)}_{xc}$ (Fig. 2b-i). 
From the further iterations we deduce the following diagrammatic rules
for $v_{xc}$ in arbitrary order: \\
{\it i}. Draw all graphs for density according to the usual rules 
\cite{LandauIX} and attach wiggled lines to external point of each graph. \\
{\it ii}. Whenever it is possible separate  the graphs into two parts by cutting 
two fermionic lines. Join the external fermionic lines of these parts and 
connect them by the wiggled line. Do not cut lines
attached to the external wiggled line. \\
{\it iii}. If a new graph is separable, repeat {\it ii}.\\
{\it iv}. If there are several possible cross sections repeat {\it ii} and
{\it iii} for every cross-section. \\
{\it v}. Leave only nonequivalent graphs.\\
For example, graphs a-e in Fig.2 appear according to 
{\it i}. Graph f is obtained from d applying {\it ii}, whereas graphs g-i
originate from the graph e by the successive application of 
{\it ii}-{\it v}. 

Given a diagrammatic representation for $v_{xc}$ we can
easily construct a graphical expansion for any quantity e.g. for one particle
Green's function, a response function or an energy. We find that
the series for the energy coincides with the energy expansion obtained
in a different 
context in Ref. \onlinecite{Jap}  (see also Ref. \onlinecite{Valiev}). 
As the diagrammatic expansion is derived maintaining the exact density
in every order,  
the series for $v_{xc}$ is in fact a graphical representation of the 
G\"orling-Levy perturbation theory (GLPT) \cite{GL}. An obvious
advantage of the  
graphical method is a possibility to construct $v^{(k)}_{xc}$ for
every $k$ in a transparent form.

An important feature of the KS-based diagram technique is that every
irreducible self-energy insertion  $\Sigma({\bf r},{\bf r}')$ is
accompanied by the {\it local}  
counter term with the opposite sign which has the structure of the
average of $\Sigma$:  
$(G_{S}\Sigma G_{S})(G_{S}G_{S})^{-1}$. This term guarantees the
density conservation 
and {\it locally} reduces the effective field $\Sigma$. The
first-order correction (Fig. 1b)  
gives an example of  this  compensation. It is interesting to note
that also the standard 
diagram technique can be reformulated in a similar fashion. One has to
explicitly  
introduce the correction to the chemical potential to compensate the 
change of the total number of particles (i.e. the averaged density) in every
order of the perturbation theory. This leads to similar, but spatially-averaged 
counter terms. However the local compensation in the KS-based technique is 
obviously more efficient. This means that KS particles are much closer to 
the true quasiparticles than bare electrons.  

Our graphical method has an obvious connection to the GW approximation
\cite{GW}. Let us collect in every order of the perturbation theory
only the bubble diagrams (e.g. the graph in Fig. 2c) and sum them up. 
The corresponding correction to the Green's function is still given by 
Fig. 1b, but with the RPA-screened interaction. While the first graph
in  Fig. 1b is  
exactly the GW self-energy, the second one deviates from the common GW
prescription.  
Instead of subtraction of the whole $v_{xc}$,  one has to use an 
{\it approximate} 
$v_{xc}$, (even if the exact $v_{xc}$ is known!). This is necessary
for the purpose 
of internal consistency and facilitates the density conservation. 
From this point of view it is clear that KS eigenvalues should describe well 
quasiparticles in metals (e.g. the shape of the Fermi surface), but  not in 
semiconductors. Indeed, for a short-ranged screened interaction in metals the
first (nonlocal) and the second (local) term in Fig. 2b almost cancel
(they would  
cancel exactly for a point interaction). Conversely, in insulators there is no 
pronounced cancellation since the interaction is long-ranged. As a
result, the correction  
to the KS energies gets larger the larger the gap is.

To study the electron-hole excitations one has to consider the linear response
function $\chi({\bf r},{\bf r}',\omega_{n})$. An integral equation for
this function
\begin{equation}
\chi(\omega_{n})={\tilde\chi(\omega_{n})} + 
{\tilde\chi(\omega_{n})}V_{C}\chi(\omega_{n}), 
\label{10}       
\end{equation}
contains irreducible polarization operator 
${\tilde\chi}(\omega_{n})$. 
We split ${\tilde\chi}(\omega_{n})$ into two parts:
${\tilde\chi}(\omega_{n})=\chi_{S}(\omega_{n}) + \Pi(\omega_{n})$,
where $\Pi(\omega_{n})$ includes all (self-energy and vertex) corrections 
to the irreducible susceptibility ${\tilde\chi}(\omega_{n})$. 
The  first-order corrections to $\chi(\omega_{n})$ are shown in
Fig. 3, where the first four terms correspond to the first-order
correction $\Pi^{(1)}(\omega_{n})$. Thus the total response function
in the first order is 
\begin{equation}
\chi(\omega_{n})\approx \chi_{S}(\omega_{n}) + \Pi^{(1)}(\omega_{n})+
\chi_{S}(\omega_{n})V_{C}\chi_{S}(\omega_{n}).
\label{11}     
\end{equation}
\begin{figure}[t]
\centering
\includegraphics[
width=0.25\textwidth]{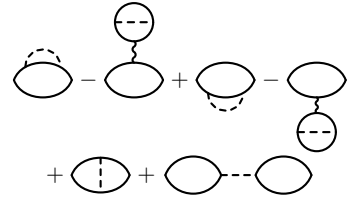}
\caption{First-order corrections to the response function.}
\label{fig3}
\end{figure}
\noindent 
The graphs in Fig. 3 display the physical meaning of all  
corrections to the electron-hole excitation energy. The
first two diagrams as well as the third and the fourth graphs are the 
self-energy corrections for an electron and a hole respectively. 
The fifth graph is the electron-hole Coulomb interaction, while the last term
is the Coulomb energy of the excitation-induced charge density.
Performing summation over frequencies and an analytic continuation in
the first five diagrams 
in Fig. 3 we obtain $\Pi^{(1)}(\omega)$ close to the excitation 
frequency $\omega^{S}_{ij}$:
\begin{eqnarray}\label{12}     
\Pi^{(1)}(\omega)&=&\gamma_{ij}\Phi_{ij}({\bf r})\Phi^{*}_{ij}({\bf r}')
\frac{\gamma_{ij}\Delta_{ij}}{\left(\omega - \omega^{S}_{ij}\right)^{2}},\\
\nonumber
\Delta_{ij}&=&\langle\psi_{i}({\bf r})|
v_{F}({\bf r},{\bf r}')-
v_{x}({\bf r})\delta({\bf r}-{\bf r}')
|\psi_{i}({\bf r}')\rangle\\
\nonumber
&-&\langle\psi_{j}({\bf r})|v_{F}({\bf r},{\bf r}')-
v_{x}({\bf r})\delta({\bf r}-{\bf r}')|\psi_{j}({\bf r}')\rangle\\
&-&\langle\Phi_{ii}({\bf r})|V_{C}({\bf r},{\bf r}')
|\Phi_{jj}({\bf r}')\rangle.
\label{13}     
\end{eqnarray}
In Eq. (\ref{13}) $v_{F}({\bf r},{\bf r}')$ is the Fock nonlocal potential
and $v_{x}({\bf r})$ is the local exchange potential or OEP. Apparently 
$\gamma_{ij}\Delta_{ij}$  is a shift of  the 
excitation energy due to the first five graphs in Fig. 3. The last
graph gives a correction which is the first term in
Eq. (\ref{5}). The total correction
$\Delta\omega^{(1)}_{ij}=
\gamma_{ij}(\Delta_{ij}+\langle\Phi_{ij}|V_{C}|\Phi_{ij}\rangle)$
coincides with the first-order 
GLPT \cite{Goerling,Gonze}. With the increase of a system volume the last
term in Eq. (\ref{13}) (the energy of the electron-hole interaction)
vanishes as $e^{2}/L$. Conversely, the first two terms remain finite
(of the order of $e^{2}/l$) and describe e.g. an exchange shift of the band gap
of a semiconductor.
 
The above results can be restated in terms of TDDFT. In the linear
approximation, i.e. after the first iteration
Eq. (\ref{1}) yields 
\begin{equation}
\chi(\omega_{n})\approx \chi_{S}(\omega_{n}) + 
\chi_{S}(\omega_{n})\left[V_{C}+ 
f_{xc}(\omega_{n})\right]\chi_{S}(\omega_{n}).
\label{14}     
\end{equation}
A comparison to Eq. (\ref{13}) leads to the following relation 
\begin{equation}
f_{xc}(\omega)=
\chi_{S}(\omega)^{-1}\Pi^{(1)}(\omega)\chi_{S}(\omega)^{-1}.
\label{15}
\end{equation}
As the linear approximation $v^{(1)}_{xc}$
is simply the x-only OEP (Fig. 2a), a linear expression
Eq. (\ref{15}) should provide a dynamical OEP kernel
$f^{OEP}_{x}(\omega)$. Indeed, a direct functional differentiation 
of the graph Fig. 2a versus density yields Eq. (\ref{15}). 

At any non-resonant frequency, e.g. in statics  $\omega=0$, the 
non-locality range of $f_{xc}$  Eq. (\ref{15}) is about an
interparticle distance $l$.   
The correction to the excitation energies Eq. (\ref{5}) depends,
however, on the kernel 
at resonance $f_{xc}(\omega^{S}_{ij})$. 
Let us consider a spatial extension of this kernel using Eq. (\ref{15}).
To calculate $\chi_{S}(\omega)^{-1}$ at resonance $\omega^{S}_{ij}$ 
one has to keep both a singular and a regular parts of
$\chi_{S}(\omega)$. Following 
Ref. \onlinecite{Gonze} we write $\chi_{S}= \gamma_{ij}\Phi_{ij}({\bf
  r})\Phi_{ij}({\bf r}')/(\omega 
-\omega^{S}_{ij}) + \chi_{r}({\bf r},{\bf r}')$, where $\chi_{r}$ is
a regular part. Substituting $\Pi^{(1)}(\omega)$ (\ref{12}) to 
Eq. (\ref{15}) and performing calculations we arrive at the following result 
\begin{eqnarray}\nonumber
& &f_{xc}({\bf r},{\bf r}',\omega_{ij})= \\
& & \Delta_{ij}\frac{\int d{\bf r}_{1}d{\bf r}_{2}
\chi^{-1}_{r}({\bf r},{\bf r}_{1})\Phi_{ij}({\bf r}_{1})
\Phi^{*}_{ij}({\bf r}_{2})\chi^{-1}_{r}({\bf r}_{2},{\bf r}')}
{\left[ \int d{\bf r}_{1}d{\bf r}_{2}\Phi^{*}_{ij}({\bf r}_{2})
\chi^{-1}_{r}({\bf r}_{2},{\bf r}_{1})
\Phi_{ij}({\bf r}_{1})\right]^{2}}.
\label{16}
\end{eqnarray}
It is seen from Eq. (\ref{16}) that a spacial scale of $f_{xc}({\bf
  r},{\bf r}',\omega_{ij})$  
is dictated by the functions $\Phi_{ij}({\bf r})$ which extend over the whole
volume. Thus the non-locality range of  the resonant $f_{xc}$ is simply a
system size, which makes possible a finite xc contribution 
$\langle\Phi_{ij}|f_{xc}(\omega_{ij})|\Phi_{ij}\rangle =
\Delta_{ij}$ in Eq. (\ref{5}). This result was considered in
Ref. \onlinecite{Gonze} as an indication of an equivalence of GLPT and
TDDFT. We emphasize that this equivalence holds only for a dynamic
 xc-kernel at resonant  
frequency and is not fulfilled by any static approximation.
    
It is straightforward  to construct a kernel $f^{GW}_{xc}(\omega)$ which 
reproduces the first-order GW result. One has to replace
$\Pi^{(1)}(\omega)$ in Eq. (\ref{15}) by the function
$\Pi_{scr}^{(1)}(\omega)$ which is defined by the first four graphs
in Fig. 3, but with the RPA-screened interaction. TDDFT formalism with this
$f^{GW}_{xc}(\omega)$  exactly reproduces the GW approach. 
For a semiconductor with the band gap $E_{g}$ the xc-kernel at 
$\omega = E_{g}$, which is responsible for the band gap correction, 
is given by Eq. (\ref{16}) with
$\Delta_{ij}=\Delta_{N+1}-\Delta_{N-1}$. Here $\Delta_{N+1}$
($\Delta_{N-1}$) are the 
discontinuities of the xc-potential upon addition (removal) of a particle.
The fundamental relation of $f_{xc}$ at resonant frequency to the 
discontinuity of $v_{xc}$ has a simple physical interpretation. The jump of
$v_{xc}$  signifies a constant shift of  a potential throughout the system due 
to addition of one particle. It means that another probe particle
interacts with the first 
one anywhere in a system. This should be interpreted as a xc-interaction
with a length scale of  the size of a system and with an amplitude equal to
the $v_{xc}$ discontinuity. Importantly, this results holds only at resonant 
frequency when a real creation of an electron-hole pair takes place.
The arguments above show that not only any static approximation, but also 
any LDA-based dynamic approximation (including any gradient corrections) 
for $f_{xc}$ cannot provide
a consistent results for excitation energies and a construction of explicit 
orbital- and frequency-dependent functionals similar to 
Eq. (\ref{15}) is required. A possible alternative is a direct calculation
of the irreducible polarization operator using the diagram method
outlined above. 
This formally allows to express excitation energies as functionals of
the KS orbitals and consequently of the ground state density.

The work of  I.T. was supported by the Alexander von Humboldt
Foundation and partly by the Russian Federal Program "Integration".

\end{multicols}
\end{document}